\documentclass[11pt,a4paper]{article}
\usepackage{graphicx}
\usepackage{dcolumn}
\usepackage{bm}
\def\nn {\nonumber}

\newcommand{\be}{\begin{equation}}
\newcommand{\ee}{\end{equation}}
\newcommand{\bea}{\begin{eqnarray}}
\newcommand{\eea}{\end{eqnarray}}

\newcommand{\Gm}{\Gamma}
\newcommand{\ep}{\epsilon}
\newcommand{\de}{\delta}

\newcommand{\om}{\omega}

\newcommand{\tht}{\theta}

\newcommand{\oD}{\overline{D}}

\newcommand{\op}{\overline\Pi}
\newcommand{\iop}{{\rm Im}\overline\Pi}
\newcommand{\ov}{\overline}

\newcommand{\vk}{\vec k}

\newcommand{\vq}{\vec q}

\newcommand{\mn}{\mu\nu}
\newcommand{\del}{\partial}

\newcommand{\F}{f_\pi}

\newcommand{\oK}{\overline{K}}

\begin{document}

\title{Effect of in-medium spectral density of $D$ and $D^*$ mesons on the $J/\psi$
dissociation in hadronic matter}
\author{Sabyasachi Ghosh$^{1,2}$, Sukanya Mitra$^2$ and Sourav Sarkar$^2$}
\date{}
\maketitle
\begin{center}
{$^1$Center for Astroparticle Physics and Space Science, 
Bose Institute, Block EN, Sector V, Salt Lake, Kolkata 700091, India}
\end{center}
\begin{center}
{$^2$Theoretical Physics Division, Variable Energy Cyclotron Centre, 
1/AF Bidhannagar, Kolkata 700064, India}
\end{center}

\begin{abstract}
The one-loop self-energy of the $D$ and $D^*$ mesons in a hot 
hadronic medium is evaluated using the real 
time formalism of thermal field theory. The interaction of the heavy open-charm
mesons with the thermalized constituents $(\pi,K,\eta)$ of the hadronic matter is 
treated in the covariant formalism of heavy meson chiral perturbation theory.
The imaginary parts are extracted from the discontinuities of 
the self-energy function across the unitary and the 
Landau cuts. The non-zero contribution from the latter
to the spectral density of $D$ and $D^*$ mesons opens a number of 
subthreshold decay channels of the $J/\psi$ leading to a significant increase
in the dissociation width in hadronic matter.
  
\end{abstract}

\maketitle
\section{Introduction}

Heavy quarks, charm and bottom, and their bound states have emerged in a leading
role as probes of the strongly interacting system produced in heavy ion
collisions~\cite{Vogt,Rapp_rev}. In fact, along with the elliptic
flow and nuclear suppression of light hadrons, these quantities have also been
measured for the single $e^-$ spectra from open charm and beauty meson decays at
RHIC~\cite{raaexpt,v2expt}. After the discovery of QGP as a near perfect fluid at
RHIC, much attention has gone into the study of its transport properties and
this is where heavy quarks are advantageous. Because of the fact that their
masses are much larger than the typical temperature even at LHC energies, they are predominantly
produced in the early primordial stages of the collision and are witness to
the entire evolution. Moreover, their thermalisation in the QGP being unlikely
they presumably retain the memory of their interaction. Their large mass
allows them to be treated as Brownian particles validating a Fokker-Planck
description for the study of transport properties using heavy quarks as probes.
This formalism has been implemented in many analyses~\cite{Ben,vanHees,Mustafa,Santosh}
and the drag and diffusion coefficients were evaluated.

In most estimates the role of hadronic matter in the
analysis of elliptic flow and nuclear suppression of heavy flavors 
was completely ignored. Recently,
the drag and diffusion coefficients 
of a hot hadronic medium consisting of pions, kaons and eta using 
$D$~\cite{Sabya_D} and $B$~\cite{Santosh_BD} mesons 
as well as their role in heavy flavour suppression~\cite{Santosh_raa} were evaluated using
interactions from heavy meson chiral perturbation theory. The significant
values obtained for these coefficients indicate a substantial interaction of the
heavy mesons with the thermal bath (see also \cite{Laine,He,Juan}). 
It is hence worthwhile to investigate how
the same interaction modifies the spectral properties of $D$ and $D^\ast$ mesons
in hot hadronic matter. This in turn should be reflected in observables like
dissociation width of the $J/\psi$ and the
dilepton spectra in the intermediate mass region where correlated $D$ meson
decays make a significant contribution~\cite{Rapp_dil}.

The suppression of the $J/\psi$ production in highly relativistic heavy ion
collisions with respect to binary scaled $pp$ collisions 
which is experimentally found
by NA50 ~\cite{NA50}, NA60 ~\cite{NA60} as well
as by the PHENIX ~\cite{Phenix} collaboration has long been suggested 
as a signal of quark-gluon plasma formation in heavy ion
collisions~\cite{Satz, Blaizot}. However, other mechanisms
based on the $J/\psi$ absorption by comoving hadrons have also been
proposed as a possible explanation~\cite{Vogt}.
In this connection the inelastic reaction rates of $J/\psi$ in the
hadronic phase have
been studied using different effective hadronic models~\cite{Haglin,Liu_Ko,Oset}.
In addition, several theoretical efforts 
~\cite{Friman_D,Hayashigaki,Kampfer,Weise_Lee,Tsushima,Gottfried,Blaschke,
Tolos1,Sibirtsev,Mishra_D,Blaschke_Jpsi,Lutz,Ramos,Molina} have 
been made in order to understand the in-medium behavior of charmed mesons 
using different hadronic models e.g. QCD sum rules
~\cite{Hayashigaki,Kampfer,Weise_Lee}, Quark-Meson Coupling (QMC)
\cite{Tsushima}, Nambu-Jona-Lasinio (NJL) model \cite{Gottfried,Blaschke}.
Most of them have predicted a larger 
mass drop of open-charm mesons
than that of $J/\psi$ which may help to explain
the $J/\psi$ suppression in a hadronic environment 
\cite{Friman_D,Hayashigaki,Sibirtsev}
although a width enhancement with negligible mass shift of open charm mesons
\cite{Tolos1,Blaschke,Blaschke_Jpsi}
have also been suggested in order to explain the suppression
\cite{Blaschke,Blaschke_Jpsi}.

In the present work we investigate in detail the in-medium modification of open
charm mesons $D$ and $D^*$
by performing an explicit evaluation of their spectral functions
in the Real-time formalism of Thermal Field theory (RTF).
The interaction vertices appearing in the one-loop self-energy of $D$ and $D^*$
is obtained from the Lagrangian 
of Chiral Perturbation Theory ($\chi PT$) involving heavy-light pseudoscalar and
vector mesons in the covariant form~\cite{Geng}.
Unlike most treatments in the literature, we follow the technique of
extracting the imaginary part from the
discontinuities of the self-energy function in the complex energy 
plane~\cite{Weldon,G_rho}.
The position of the branch cuts, the Landau cut in particular, which appears only
in the medium along with the usual unitary cut which also exists in vacuum 
but is now weighted by the phase space distributions, are determined from the 
kinematics before evaluating their relative contributions.
Using the in-medium spectral function of  $D$ and $D^*$ mesons we then 
investigate its effect on the $J/\psi$ dissociation in hadronic matter by 
calculating its width in the different $D\oD$, 
$D\overline{D}^*$ and $D^*\overline{D}^*$ channels.

The manuscript is organized as follows.
Section II deals with the self-energy of $D$ and $D^*$ mesons and
 begins with a discussion of the structure of 
the propagator in the RTF. After a  
discussion of the interaction 
vertices obtained from the chiral Lagrangian, the branch cut structure of 
the one-loop self-energy is elaborated upon. 
The results are presented in section III,
followed by a summary in section IV.

\section{The self-energy of the $D$ and $D^\ast$}

\subsection{Structure of interacting propagator in RTF}
In the real time formalism of thermal field theory~\cite{Bellac}, all two point 
functions have 
a (2$\times$2) matrix structure. These matrices may however be diagonalized in terms 
of a single analytic function that determines completely the dynamics of the 
corresponding two point function~\cite{Kobes,Mallik}. Since this function is related simply to any 
one component, say the 11 component of the matrix, we need to calculate only this 
component of the matrix.

The 11 component of the free thermal matrix propagator has two parts. 
Along with the usual vacuum term there is a part containing 
the on-shell distribution functions of like particles in the medium. 
Here pseudoscalar and vector mesons will appear in our calculation 
and hence we discuss their 
11 components of the propagators in the following. For the pseudoscalar,
this is given by 
\begin{equation}
G^{(0)11}(q)=\Delta^{(0)}(q)+2\pi i n\delta(q^2-m_q^2)
\label{D11}
\end{equation}
where $\Delta^{(0)}$ is the vacuum propagator of the scalar field and $n$ its 
equilibrium distribution function given by
\be
\Delta^{(0)}(q)=\frac{-1}{q^2-m_q^2+i\eta},~~~~n(\om_q)=\frac{1}{e^{\beta\om_q}-1}
\ee
with $~~\om_q=\sqrt{\vec q^2+m_q^2}$.

The Dyson equation relates 
the complete propagator (matrix) $G^{ab}$ 
with the free propagator $G^{(0)ab}$ 
and self-energy $\Pi^{ab}$ matrices at finite temperature,
\begin{equation}
G^{ab}=G^{(0)ab}-G^{(0)ac}\Pi^{cd} G^{db}~.
\end{equation}
where $a, b, c, d$ are thermal indices and take the values 1 and
2~\cite{Bellac}. 
These matrices can be diagonalized in terms of their respective analytic
 functions (denoted by a bar) which again obey
\begin{equation}
{\overline G}={\overline G^{(0)}}-{\overline G^{(0)}}{\overline{\Pi}}~ {\overline G}
~,~{\rm where}~{\overline G^{(0)}}=\Delta^{(0)}(q)
\end{equation} 
and can be solved to get the full propagator $\overline G$. 
The diagonal element of self-energy function is related to the 11 component 
as~\cite{Kobes,Mallik},
\begin{eqnarray}
&&~~~~~~~~~\mbox{Re}{\overline\Pi}(q)=\mbox{Re} \Pi_{11}(q)
\nonumber\\
&&\mbox{Im} {\overline\Pi}(q) =\epsilon(q_0)\tanh(\frac{\beta q^{0}{2}})\mbox{Im} \Pi_{11}(q)
\label{re_im_bar}
\end{eqnarray}
in terms of which the spectral function of the $D$ meson is defined as
\be
A_{D}=2\ep(q_0){\rm Im}\overline G(q)
\ee
where
\begin{equation}
{\rm Im}\overline G(q)=\frac{-\sum \mbox{Im}\overline\Pi}{(q^2-m^2_D-\sum 
{\rm Re}\overline\Pi)^2+(\sum \mbox{Im}\overline\Pi)^2}~,
\label{spec}
\end{equation}
the summation running over all the loops. To obtain the in-medium spectral
function one thus evaluates the real and imaginary parts of the 11-component of
the self-energy.

The $D^\ast$ being a vector meson, the propagator contains tensor indices.
The 11-component in this case is given by, 
\begin{equation}
G^{(0)11}_{\mu\nu}(p)=(-g_{\mu\nu}+\frac{p_{\mu}p_{\nu}}{m_p^2})G^{(0)11}(p)~.
\end{equation}
As shown in~\cite{Kobes,G_rho_dense} the full propagator can be obtained
analogously as above.
The spectral function is then given by eq.~(\ref{spec}) above with
the spin-averaged self-energy 
\begin{equation}
\overline\Pi=\frac{1}{3} \overline\Pi^{\mu}_{\mu}~.
\label{Dstar}
\end{equation}

\subsection{Interaction vertices}

\begin{figure}[ht]
\begin{center}
\includegraphics[scale=0.6]{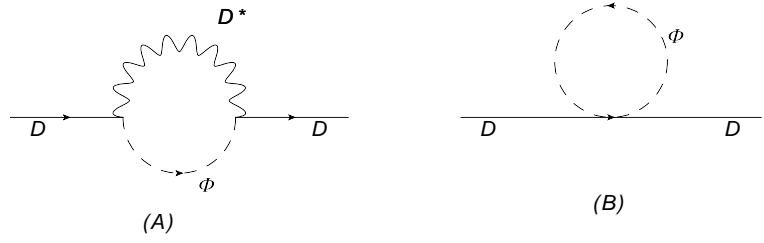}
\caption{One loop diagram of D meson self-energy where
$ \Phi$ stands for $\pi,\eta$ and $K$ mesons
}
\label{fig1}
\end{center}
\end{figure}
\begin{figure}[ht]
\begin{center}
\includegraphics[scale=0.8]{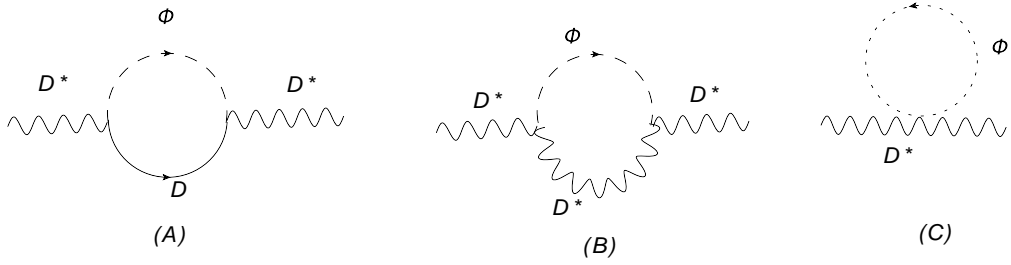}
\caption{One loop diagram of $D^*$ meson self-energy}
\label{Dst_loop}
\end{center}
\end{figure}

The vertices appearing in the one-loop self-energy graphs 
are obtained using chiral perturbation theory. 
The lowest order chiral Lagrangian for the heavy-light pseudoscalar 
and vector meson is~\cite{Geng},
\bea
\mathcal{L}&=&\langle \mathcal{D}_\mu P \mathcal{D}^\mu P^\dagger\rangle
-m_D^2\langle PP^\dagger\rangle 
-\langle \mathcal{D}_\mu P^{*\nu} \mathcal{D}^\mu P^{*\dagger}_\nu\rangle +
 m_{D}^2\langle P^{*\nu} P^{*\dagger}_\nu\rangle
\nn\\
&&+ ig \langle P^*_\mu u^\mu P^\dagger -P u^\mu P^{*\dagger}_\mu\rangle 
+\frac{g}{2m_D}\langle(P^*_\mu u_\alpha\del_\beta P^{*\dagger}_\nu -\del_\beta P^*_\mu u_\alpha P^{*\dagger}_\nu)
\epsilon^{\mu\nu\alpha\beta}\rangle
\label{lag:lo}
\eea
where $P=(D^0,D^+,D^+_s)$ and $P^*_{\mu}=(D^{*0},D^{*+},D^{*+}_s)_\mu$ are the 
triplets of $D$ and $D^*$ meson fields.
The covariant derivatives are defined as $\mathcal{D}_\mu P_a=\partial_\mu P_a  - \Gamma_\mu^{ba} P_b$ 
and $\mathcal{D}^\mu P^\dagger_a=\partial^\mu P^\dagger_a+\Gamma^\mu_{ab} P^\dagger_b$, with
$a,b$ the $SU(3)$ flavor indices. $m_D$ is the mass of heavy-light meson 
in the chiral limit. The value of the heavy-light pseudoscalar-vector
coupling constant $g=1.177$ GeV is obtained by reproducing the
experimental $D^*\rightarrow D\pi$ decay width of $\sim$ 65 keV
with the above interaction\cite{Geng}.
The vector and axial-vector currents are respectively given by
\bea
\Gamma_\mu &=&\frac{1}{2}(u^\dagger\partial_\mu u+u\partial_\mu u^\dagger)
\nn\\
\quad u_\mu &=& i(u^\dagger\partial_\mu u-u\partial_\mu u^\dagger)
\eea
where $u={\rm exp}(\frac{i\lambda_a\Phi_a}{2\F})$ collects the octet
of Nambu Goldstone fields with
\be
 \lambda_a\Phi_a=\sqrt{2}\left(
\begin{array}{ccc}
 \frac{\pi^0}{\sqrt{2}}+\frac{\eta}{\sqrt{6}} & \pi^+ &K^+\\
  \pi^- & -\frac{\pi^0}{\sqrt{2}}+\frac{\eta}{\sqrt{6}} & K^0\\
  K^- & \bar{K}^0 & -\frac{2}{\sqrt{6}}\eta
\end{array}
\right)
\label{lag}
\ee
and $\F$ is the pseudoscalar decay constant in the chiral limit.
To lowest order in $\Phi$ the vector and axial-vector currents are
\be
\Gm_\mu=\frac{1}{8\F^2} [\Phi , \del_\mu \Phi], ~~~~~u_\mu=-\frac{1}{\F}
\del_\mu \Phi
\ee
from which the interaction terms are separately obtained as
\be
\mathcal{L}_{P\Phi P\Phi}=\frac{1}{8\F^2}\langle \partial_\mu P [\Phi,\partial^\mu\Phi] P^\dagger 
-P [\Phi,\partial^\mu\Phi]\partial_\mu P^\dagger \rangle
\label{4_line}
\ee
\be
\mathcal{L}_{P^*\Phi P^*\Phi}=-\frac{1}{8\F^2}\langle \partial_\mu P^{*\nu}[\Phi,\partial^\mu\Phi]
P^{*\dagger}_\nu -P^{*\nu}[\Phi,\partial^\mu\Phi]\partial_\mu P^{*\dagger}_\nu\rangle
\label{4_line_Dst}
\ee
\be
\mathcal{L}_{P^*P\Phi}=-i\frac{g}{\F} \langle P^*_\mu \partial^\mu\Phi P^\dagger 
-P \partial^\mu\Phi P^{*\dagger}_\mu\rangle
\label{3_line}
\ee
\be
\mathcal{L}_{P^*P^*\Phi}=-\frac{g}{2m_D\F}\langle(P^*_\mu \del_\alpha\Phi\del_\beta P^{*\dagger}_\nu 
-\del_\beta P^*_\mu \del_\alpha\Phi P^{*\dagger}_\nu)
\epsilon^{\mu\nu\alpha\beta}\rangle 
\label{3_line_Dst}
\ee
Considering e.g.
the $D^+$ from the triplet $P$, the relevant
part of Eq.(\ref{4_line}) and (\ref{3_line}) are respectively
given by
\bea
&&\mathcal{L}_{D^+\Phi D^+\Phi}=\frac{-1}{4\F^2} [ (D^+\partial_\mu\pi^+\partial^\mu D^-\pi^-
-\partial^\mu D^+\partial_\mu\pi^+D^-\pi^- +\partial^\mu D^+\pi^+D^-\partial_\mu\pi^-
- D^+\pi^+\partial^\mu D^-\partial_\mu\pi^- )
\nonumber\\
&&~~~~~~~~~~~~~+(D^+\partial_\mu\oK^0\partial^\mu D^-K^0 
- \partial^\mu D^+\partial_\mu\oK^0D^-K^0 +\partial^\mu D^+\oK^0D^-\partial_\mu K^0
- D^+\oK^0\partial^\mu D^-\partial_\mu K^0 )],
\label{4_line2}
\eea
\bea
\mathcal{L}_{D^+P^*\Phi}&=&i\frac{g}{\F} 
[\sqrt{2}(D^+\partial^\mu\pi^-\oD^{*0}_{\mu}-D^-\partial^\mu\pi^+D^{*0}_{\mu})
+(D^-\partial^\mu\pi^0D^{*+}_{\mu}-D^+\partial^\mu\pi^0D^{*-}_{\mu}) 
\nonumber\\
&&+\frac{1}{\sqrt{3}}(D^+\partial^\mu\eta D^{*-}_{\mu} -D^-\partial^\mu\eta D^{*+}_{\mu}) 
+\sqrt{2}(D^+\partial^\mu K^0D^{*-}_{s\mu}-D^-\partial^\mu\overline{K}^0D^{*+}_{s\mu})]
\label{3_line2}~.
\eea
Similarly for $D^{*+}$ from the triplet $P^*$ the relevant
part of Eq.(\ref{4_line_Dst}), (\ref{3_line}) and (\ref{3_line_Dst}) are respectively
given by
\bea
\mathcal{L}_{D^{*+}\Phi D^{*+}\Phi}&=&\frac{1}{4\F^2} [(D^{*+\nu}\partial_\mu\pi^+\partial^\mu D^{*-}_\nu\pi^-
-\partial^\mu D^{*+\nu}\partial_\mu\pi^+D^{*-}_\nu\pi^- +\partial^\mu D^{*+\nu}\pi^+D^{*-}_\nu\partial_\mu\pi^-
\nn\\
&&- D^{*+\nu}\pi^+\partial^\mu D^{*-}_\nu\partial_\mu\pi^- )
+(D^{*+\nu}\partial_\mu\oK^0\partial^\mu D^{*-}_\nu K^0 
- \partial^\mu D^{*+\nu}\partial_\mu\oK^0D^{*-}_\nu K^0
\nn\\
&&~~~~~~~~~+\partial^\mu D^{*+\nu}\oK^0D^{*-}_\nu\partial_\mu K^0
- D^{*+\nu}\oK^0\partial^\mu D^{*-}_\nu\partial_\mu K^0 )]~,
\label{4_line_Dst2}
\eea
\bea
\mathcal{L}_{D^{*+}P\Phi}&=&i\frac{g}{\F} 
[\sqrt{2}(D^+\partial^\mu\pi^-\oD^{*0}_{\mu}-D^-\partial^\mu\pi^+D^{*0}_{\mu})
+\sqrt{2}(D^-\partial^\mu\pi^0D^{*+}_{\mu}-D^+\partial^\mu\pi^0D^{*-}_{\mu})
\nonumber\\
&&+\frac{1}{\sqrt{3}}(D^+\partial^\mu\eta D^{*-}_{\mu} -D^-\partial^\mu\eta D^{*+}_{\mu}) 
+\sqrt{2}(D^+_s\partial^\mu \ov K^0D^{*-}_{\mu}-D^-_s\partial^\mu{K}^0D^{*+}_{\mu})],
\label{3_line2_Dst}
\eea
\bea
\mathcal{L}_{D^{*+}P^*\Phi}&=&\frac{g\ep^{\mu\nu\alpha\beta}}{2m_D\F} 
[\sqrt{2}(\del_\beta D^{*+}_\mu\partial_\alpha\pi^-\oD^{*0}_{\nu}
-D^{*0}_{\mu}\partial_\alpha\pi^+\del_\beta D^{*-}_\nu)
+(D^{*+}_{\mu}\partial_\alpha\pi^0\del_\beta D^{*-}_\nu
-\del_\beta D^{*+}_\mu\partial_\alpha\pi^0D^{*-}_{\nu}) 
\nonumber\\
&&+\frac{1}{\sqrt{3}}(\del_\beta D^{*+}_\mu\partial_\alpha\eta D^{*-}_{\nu} 
-D^{*+}_{\mu}\partial_\alpha\eta\del_\beta D^{*-}_\nu) 
+\sqrt{2}(\del_\beta D^{*+}_\mu\partial_\alpha K^0D^{*-}_{s\nu}
-D^{*+}_{s\mu}\partial_\alpha\overline{K}^0\del_\beta D^{*-}_{\nu})]
\label{3_line_Dst2}
\eea
Having fixed the propagator and interaction vertices we are now in a position to
write down the expression for the self-energy function. 
The expression of $D^+$ meson self-energy coming from
fig.\ref{fig1}(A) and (B) are respectively given by  
\begin{equation}
\Pi^{11}(q)=i\int \frac{d^4k}{(2\pi)^4}N(q,k)D^{11}(k,m_k)D^{11}(q-k,m_p)
\label{self_3line}
\end{equation}
and
\begin{equation}
\Pi^{11}(q)=i\int \frac{d^4k}{(2\pi)^4}N(q,k)D^{11}(k,m_k)~.
\label{self_4line}
\end{equation}
The $D^{*+}$ being a vector meson its self-energy will have tensor indices.
The corresponding expressions for fig.~\ref{Dst_loop}(A), \ref{Dst_loop}(B)
and  \ref{Dst_loop}(C)
are given by Eqs.~(\ref{self_3line}) and (\ref{self_4line})
with $N(q,k)\to N^{\mu\nu}(q,k)$ and
consequently $\Pi^{11}(q)\to\Pi^{\mu\nu\,11}(q) $. The term $N(q,k)$ (and 
$N^{\mu\nu}(q,k)$) results from factors coming from the vertices and 
propagators in the self-energy diagrams.

We have considered four possible loops $\pi^{+}D^{*0}$, $\pi^{0} D^{*+}$, $\eta D^{*+}$ 
and $\overline{K}^{0} D^{*+}_{s}$ to evaluate the $D^{+}$ and $D^{*+}$ meson 
self-energy in the medium which 
are represented in fig.\ref{fig1}(A) and fig.\ref{Dst_loop}(B) respectively.
They appear due to the interactions in (\ref{3_line2}) and (\ref{3_line_Dst2})
respectively.
The $\pi^{+}D^{0}$, $\pi^{0}D^{+}$, 
$\eta D^{+}$ and ${K}^{0}D^{+}_{s}$ loops additionally appear for the
$D^{*+}$ meson self-energy due to the interactions in (\ref{3_line2_Dst})
which are shown in fig.\ref{Dst_loop}(A).
The form of $N(q,k)$ and $N^{\mu\nu}(q,k)$ in the expression 
of $D$ and $D^*$ meson self-energy are respectively obtained as
\be
N(k,q)=-\alpha^2(\frac{g}{\F})^2[k^2-\frac{(k\cdot q - k^2)^2}{m_p^2}]~
[{\rm for~fig.\ref{fig1}(A),~ from~ Eq.(\ref{3_line2})}]
\label{N_D}
\ee
\be
N^{\mu\nu}(k,q)=-\alpha^2(\frac{g}{\F})^2[k^{\mu}k^{\nu}]
~[{\rm for~fig.\ref{Dst_loop}(A),~ from~ Eq.(\ref{3_line2_Dst})}]
\ee
\bea
N^{\mu\nu}(k,q)&=&-\alpha^2(\frac{g}{m_D\F})^2[k^2q^2A^{\mn}+B^{\mn}]
~[{\rm for~fig.\ref{Dst_loop}(B),~ from~ Eq.(\ref{3_line_Dst2})}]
\nn\\
{\rm where}~A^{\mn}&=&-g^{\mn}+\frac{q^\mu q^\nu}{q^2}~,
\nn\\
B^{\mn}&=&q^2k^\mu k^\nu -(q\cdot k)(q^\mu k^\nu +q^\nu k^\mu)+(q\cdot k)^2g^{\mn}~.
\eea
In above equations, $\alpha=\sqrt{2},1,\sqrt{2},\frac{1}{\sqrt{3}}$ for 
$\pi^{+} D^{*0}$ ($\pi^{+} D^{0}$), $\pi^{0} D^{*+}$ ($\pi^{0} D^{+}$),
$K^{0} D^{*+}_{s}$ ($K^{0} D^{+}_{s}$) and $\eta D^{*+}$ ($\eta D^{+}$) 
loops respectively and $m_p$ in Eq.(\ref{N_D}) is the mass of $D^*$ meson.
Using the interaction Lagrangian (\ref{4_line2}) for $D^+$ and (\ref{4_line_Dst2}) 
for $D^{*+}$, the integrands in Eq.(\ref{self_4line}) turn out to be odd 
functions of the integration variable $k$. 
Therefore fig.\ref{fig1}(B) and fig.\ref{Dst_loop}(C) do not contribute to the
spectral function.

\subsection{Branch cut structure of the self energy}
The integral over $k_0$ is easily performed by choosing suitable contours.
The imaginary part of the self-energy function is then obtained from
the 11-component using (\ref{re_im_bar}) to get 
\bea
&&\iop(q)=-\pi\int\frac{d^3\vec k}{(2\pi)^3 4\om_k\om_p}\times
\nonumber\\
&&[N(k_0=\om_k)\{(1+n(\om_k)+n(\om_p))\de(q_0-\om_k-\om_p)
\nonumber\\
&&~~~-(n(\om_k)-n(\om_p))\de(q_0-\om_k+\om_p)\}
\nonumber\\
&& +N(k_0=-\om_k)\{(n(\om_k)-n(\om_p))\de(q_0+\om_k-\om_p)
\nonumber\\
&&~~~-(1+n(\om_k)+n(\om_p))\de(q_0+\om_k+\om_p)\}]
\label{im_cut}
\eea
where $\om_k =\sqrt{\vk^2+m_k^2}$ and $\om_p =\sqrt{(\vq-\vk)^2+m_p^2}$ 
are the energies of light pseudoscalar mesons
and the heavy-light charm mesons respectively.
The cuts in the self-energy function correspond to regions in which the four
terms in the above expression are non-vanishing and are determined by the
respective $\de$-functions. A little inspection suggests that 
the first and the fourth terms are finite for
$q^2\ge (m_p+m_k)^2$ giving rise to the unitary cut whereas the second 
and third terms are non-vanishing for $q^2\le(m_p-m_k)^2$
producing the Landau cut. The unitary cut arises as a consequence of gain
or loss of $D$  (or $D^\ast$) mesons due to formation from or decay into the
particles in the medium.
The Landau type of discontinuity appears due to disappearance
of the heavy mesons as a result of scattering with the bath particles.
While the unitary cut exists in the vacuum also, the Landau cut is purely a
medium effect.

The thermal contribution to the real part is given by
\bea
&&{\rm Re}\,\op (q)={\cal P}\int\frac{d^3\vk}{(2\pi)^3}[\frac{n(\om_k)}{2\om_k}\{
\frac{N(k_0=\om_k)}{(q_0-\om_k)^2-{\om_p}^2}
\nonumber\\
&&+\frac{N(k_0=-\om_k)}{(q_0+\om_k)^2-{\om_k}^2}\}
+\frac{n(\om_p)}{2\om_p}\{ \frac{N(k_0=q_0-\om_p)}{(q_0-\om_p)^2-\om_k^2}
\nonumber\\
&&~~~~~~~~~~~~~~~+\frac{N(k_0=q_0+\om_p)}{(q_0+\om_p)^2-\om_k^2}\}]
\label{real}
\eea
where ${\cal P}$ indicates the principal value.
The vacuum contribution is divergent and is assumed to renormalize the mass
of the $D$ mesons to their physical values. 

Writing $d^3\vk=2\pi\sqrt{\om_k^2-m_\pi^2}\,\om_k d\om_k\,\sin\tht d\tht$, where $\tht$
is the angle between $\vq$ and $\vk$, we can readily integrate over $\cos \theta$ 
using the $\de$-functions yielding 
\be
\cos \tht_0 =
\frac{-R^2 \pm 2q_0\om_k}{2|\vq|\sqrt{\om_k^2-m_k^2}}\,,~~~~~R^2=q^2-m_p^2+m_k^2\,.
\ee
with '$+$' the for first two terms and '$-$' for the last two terms of Eq.~(\ref{im_cut}).

In the present study we will be interested in the region 
$q^2>0$ and $q_0>0$.
The imaginary parts are given by~\cite{G_rho}
\bea
&&\iop=-\frac{1}{16\pi |\vq|}\int_{{\om'_-}}^{{\om'_+}} d\om_k N(k_0=-\om_k)
\nonumber\\
&&\{n(\om_k)-n(q_0+\om_k)\}
\label{im_L}
\eea
for $|\vq|\leq q_0\leq \sqrt{(m_p-m_k)^2 +|\vq|^2}$ (region of Landau cut)
and
\bea
&&\iop=-\frac{1}{16\pi |\vq|}\int_{\om_-}^{\om_+} d\om N(k_0=\om_k)
\nonumber\\
&&\{1+n(\om_k)+n(q_0-\om_k)\}
\label{im_U}
\eea
for $q_0\geq \sqrt{(m_p+m_k)^2 +|\vq|^2}$ (region of unitary cut)
with
$\om_{\pm}=\frac{R^2}{2q^2}(q_0\pm |\vq|v)\,,~~~
{\om'_{\pm}}=\frac{R^2}{2q^2}(-q_0\mp |\vq|v)$

and $v(q^2)=\sqrt{1 -\frac{4q^2 m_k ^2}{R^4}}$.

\begin{figure}
\begin{center}
\includegraphics[scale=0.32]{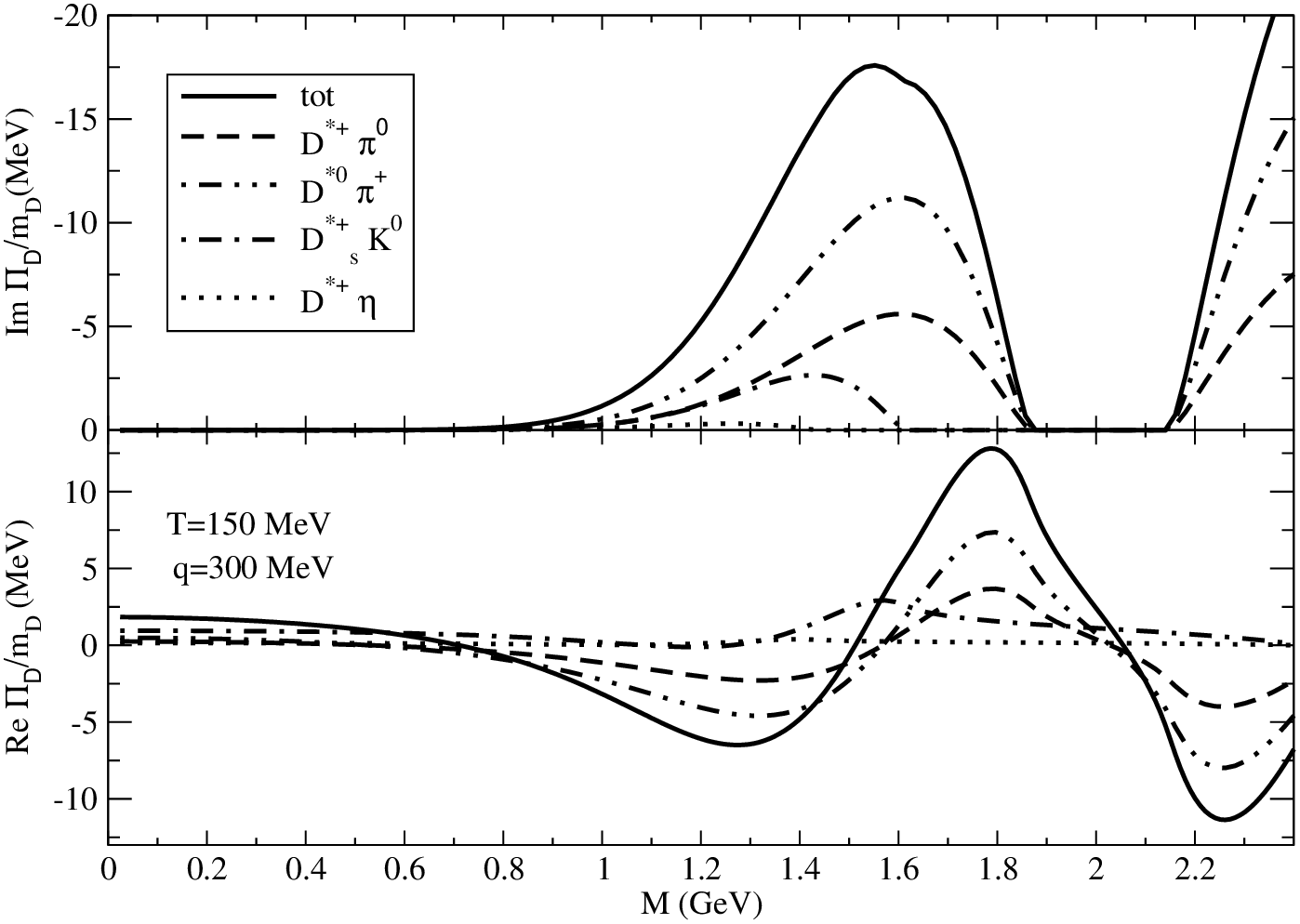}
\caption{Imaginary and real part of $D$ meson self-energy for different 
$D^*\Phi$ loops in the upper and lower panel respectively.}
\label{fig2}
\end{center}
\end{figure}
\begin{figure}
\begin{center}
\includegraphics[scale=0.32]{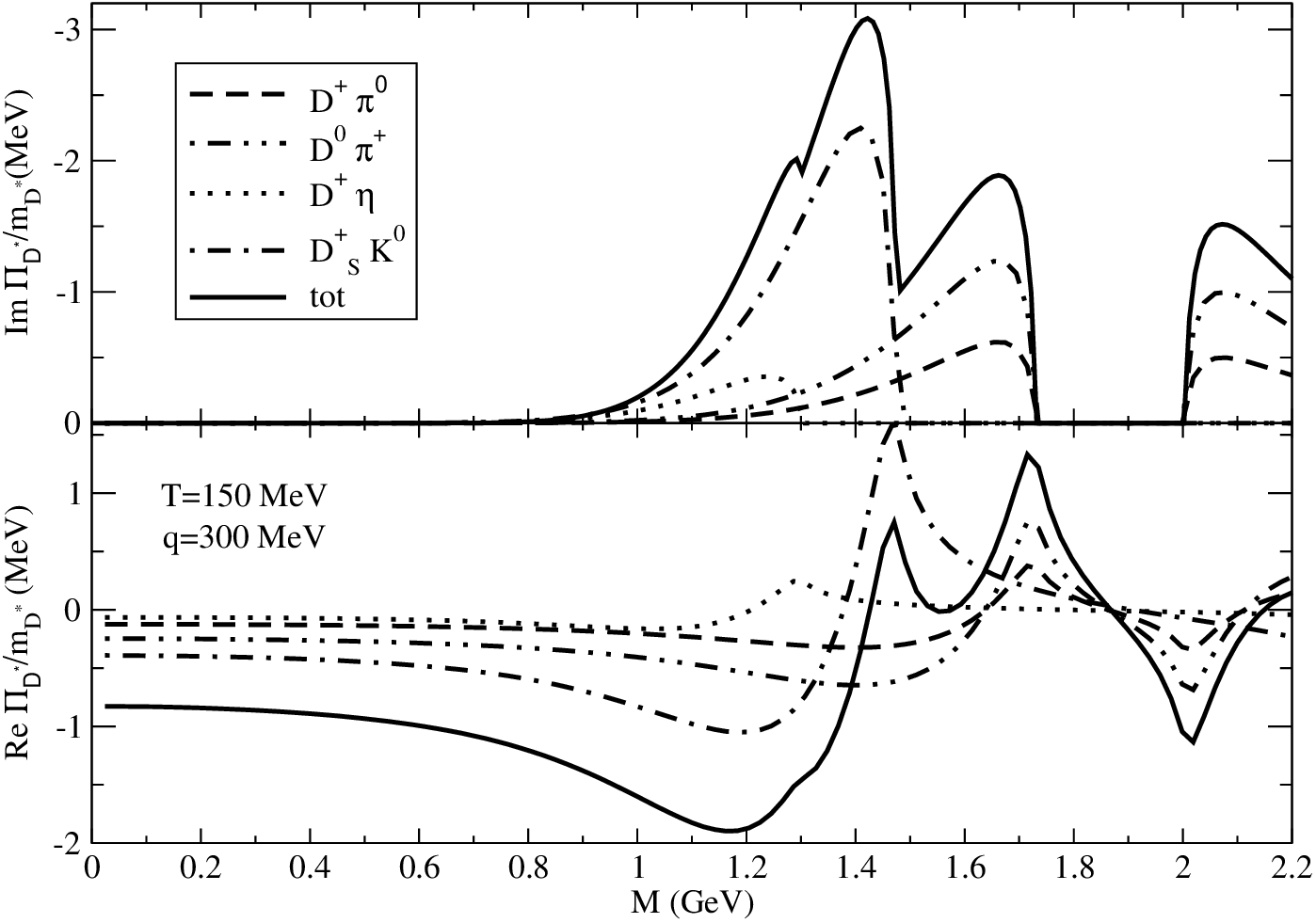}
\caption{Imaginary and real part of $D^*$ meson self-energy for different 
$D\Phi$ loops in the upper and lower panel respectively.}
\label{fig3}
\end{center}
\end{figure}
\begin{figure}
\begin{center}
\includegraphics[scale=0.32]{DstDstFi.eps}
\caption{Imaginary and real part of $D^*$ meson self-energy for different 
$D^*\Phi$ loops in the upper and lower panel respectively.}
\label{DstFi_loop}
\end{center}
\end{figure}
\begin{figure}
\begin{center}
\includegraphics[scale=0.32]{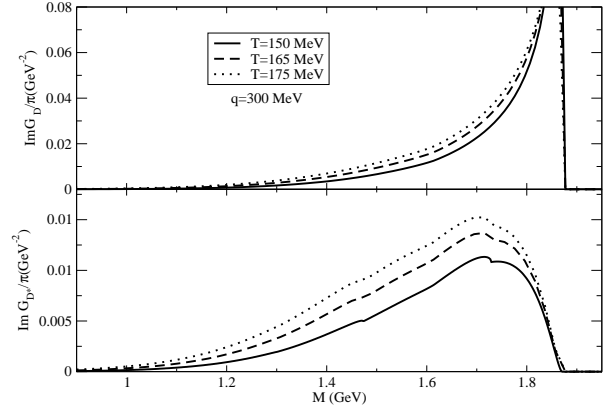}
\caption{Landau part of spectral function of $D$ and $D^*$ in the upper and 
lower panel respectively for three different temperatures.
}
\label{fig5}
\end{center}
\end{figure}
\begin{figure}
\begin{center}
\includegraphics[scale=0.32]{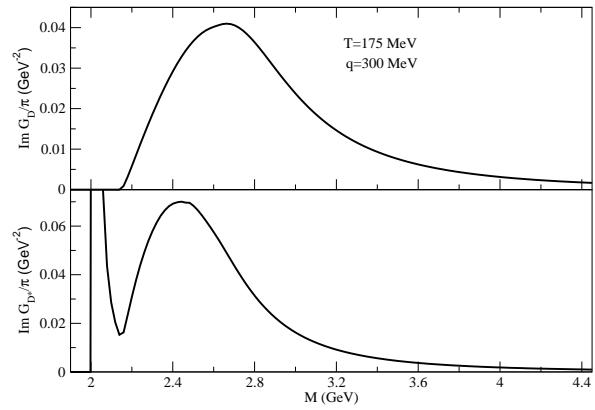}
\caption{The unitary part of the spectral functions of $D$ and $D^*$ mesons
in the upper and lower panels respectively at $T$ = 175 MeV.}
\label{fig6}
\end{center}
\end{figure}

\section{Results and Discussion}

We now present results of numerical calculation. 
For the $D$ meson $m_k=m_{\pi},m_{\eta},m_{K^0}$
and $m_p=m_{D^*},m_{D^*_s}$ whereas for the $D^*$
meson $m_p=m_{D},m_{D_s}$. 
The $\iop$ (upper panels) and Re$\op$ (lower panels) of $D$ meson 
for $D^*\Phi$ loops is shown in fig.~(\ref{fig2}) whereas
same of $D^*$ meson are shown in fig.~(\ref{fig3}) and (\ref{DstFi_loop})
for $D\Phi$ and $D^*\Phi$ loops respectively.
In the upper panels of both curves,the branch cut 
structure of $\op(q^0,\vq)$ is clearly seen. The Landau cut
contributes in the region $0\le M\le (m_p-m_k)$ and the unitary cut 
for $M\ge(m_p+m_k)$ thus
leaving in-between a region 
where the imaginary part is exactly zero.
It is easy to see that for the $D$ meson which has a nominal mass of 1867 MeV,
only the Landau cut of the $D^*\pi$ loop contributes at the pole
whereas for the $D^*$ meson the the unitary cut of the $D\pi$ loop contributes 
at the $D^*$ pole (2008 MeV). The real part however receives contribution from
all the four terms in Eq.~(\ref{real}) as seen in the lower panels of 
all these figures.

We now present the results for the spectral function of the $D$ and $D^\ast$
mesons. For the vector case we take a spin average of the imaginary part of the
full propagator as done for the self-energy in eq.~(\ref{Dstar}). Since the
entire spectral shape is of importance and not only its value at the pole, we
show the contributions from the Landau and unitary parts separately in 
figs.~(\ref{fig5}) and (\ref{fig6}) respectively. As discussed previously
the strength in the Landau region stems from
processes in which the $D$ (upper panel of fig.~(\ref{fig5})) and
$D^\ast$ (lower panel of fig.~\ref{fig5}) mesons are lost or gained
in the medium as a result of inelastic collisions with the thermalised light
mesons $\pi$, $K$ and $\eta$. Analogously decay and regeneration processes
involving the $D$ and $D^\ast$ mesons leads to the strength in the unitary
region as shown in fig.~(\ref{fig6}).
\begin{figure}
\begin{center}
\includegraphics[scale=0.32]{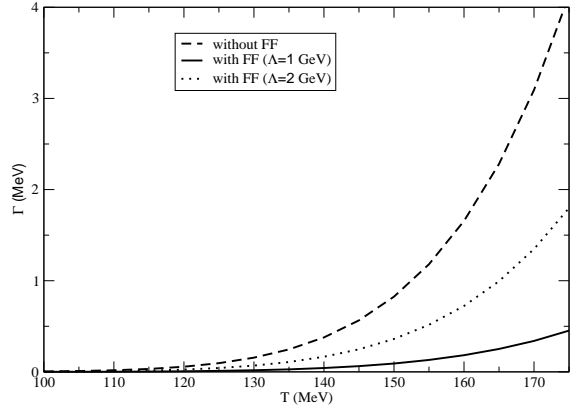}
\caption{The total dissociation width of
$J/\psi$  plotted against temperature.}
\label{fig7}
\end{center}
\end{figure}

The observed multiple structures in the spectral function have a non-trivial
effect on the $J/\psi$ dissociation in hadronic matter. For this purpose we
treat the $D$ and $D^\ast$ mesons mesons as resonances with temperature
dependent finite collisional widths as described by their spectral densities.
With the help of the phenomenological Lagrangian densities~\cite{Krein,Liu_Ko}
\bea
{\cal L}_{\psi DD}&=&ig_{\psi}\psi^\mu[\overline{D}(\del_\mu D)-(\del_\mu \overline{D})D],
\nn\\
{\cal L}_{\psi DD^*}&=&\frac{g_{\psi}}{m_\psi}\epsilon_{\alpha\beta\mu\nu}(\del^\alpha\psi^\beta)
[(\del^\mu\overline{D}^{*\nu})D+\overline{D}(\del^\mu D^{*\nu})],
\nn\\
{\cal L}_{\psi D^*D^*}&=&ig_{\psi}\{\psi^\mu[(\del_\mu\overline{D}^{*\nu})D_{*\nu}
-\overline{D}^{*\nu}(\del_\mu D_{*\nu})]+
[(\del_\mu\psi^\nu)\overline{D}_{*\nu}
\nn\\
&&~~-\psi^\nu(\del_\mu \overline{D}_{*\nu})]D^{*\mu}+
{\overline D}^{*\mu}[\psi^\nu(\del_\mu D_{*\nu})
-(\del_\mu\psi^\nu)D_{*\nu}]\}
\eea 
the expression for the in-medium 
$J/\psi$ dissociation width for $D\overline{D}$, $D\overline{D}^*$ and 
$D^*\overline{D}^*$ channels are given by,
\bea
\Gamma_{\rm med}(J/\psi \rightarrow D\oD)&=&\int\frac{g_{\psi}^2F^2}{3\pi m^2_{\psi}}
|\vec {P}_{cm}(p^2_D,p^2_{\oD})|^3
\frac{{\rm Im} G_{D}(p^{0}_{D},~\vec {P}_{cm})}{\pi}~
\frac{{\rm Im} G_{\oD}(p^0_{\oD},~\vec {P}_{cm})}{\pi}~ dp^{2}_{D}~ dp^{2}_{\oD}
\nn\\
\Gamma_{\rm med}(J/\psi \rightarrow D\oD^*)&=&\int\frac{g_{\psi}^2F^2}{3\pi m^2_{\psi}}
|\vec {P}_{cm}(p^2_D,p^2_{\oD^*})|^3
\frac{{\rm Im}G_{D}(p^{0}_{D},~\vec {P}_{cm})}{\pi}~
\frac{{\rm Im}G_{\oD^*}(p^0_{\oD^*},~\vec {P}_{cm})}{\pi}~dp^{2}_{D}~dp^{2}_{\oD^*}
\nn\\
\Gamma_{\rm med}(J/\psi \rightarrow D^*\oD^*)&=&\int\frac{g_{\psi}^2F^2}{3\pi m^2_{\psi}}
|\vec {P}_{cm}(p^2_{D^*},p^2_{\oD^*})|^3~[\frac{m^4_{\psi}+10m^2_{\psi}(p^2_{D^*}
+p^2_{\oD^*})+p^4_{D^*}+p^4_{\oD^*}+10p^2_{D^*}p^2_{\oD^*}}{4p^2_{D^*}p^2_{\oD^*}}]
\nn\\
&&~~~\times\frac{{\rm Im} G_{D^*}(p^{0}_{D^*},~\vec {P}_{cm})}{\pi}
\frac{{\rm Im} G_{\oD^*}(p^0_{\oD^*},~\vec {P}_{cm})}{\pi}~ dp^{2}_{D^*}~ dp^{2}_{\oD^*}
\label{J}
\eea
where $\vec {P}_{cm}=\frac{\sqrt{[m^2_{\psi}-(p_{D}+p_{\oD})^2]
[m^2_{\psi}-(p_{D}-p_{\oD})^2]}}{2m_{\psi}}$ 
is the center of mass momentum   
and $p_{D}=\sqrt{(p_D^{0})^2-\vec {P}^2_{cm}}$, $p_{\oD}=\sqrt{(p_{\oD}^{0})^2-\vec {P}^2_{cm}}$ 
are the invariant masses of $D$ and $\oD$ respectively. 
We take $g_{\psi}=7.8$ as in~\cite{Friman_D}.
After integrating over $p^{2}_{D}$ and 
$p^{2}_{\oD}$ the total $J/\psi$ dissociation width for all
the channels ($D\overline{D}$, $D\overline{D}^*$ as well as 
$D^*\overline{D}^*$) is shown in fig.(\ref{fig7}) as
a function of temperature. In order to effectively account for the
composite nature of the hadrons involved we incorporate a form factor,  
$F=\frac{\Lambda^2}{\Lambda^2+\vec {P}_{cm}^2}$ at the vertex
and show results for two values of the cut-off $\Lambda$.
The solid line for which  $\Lambda$=1 GeV
appears to be in reasonable agreement with~\cite{Fuch}.
\begin{figure}
\begin{center}
\includegraphics[scale=0.32]{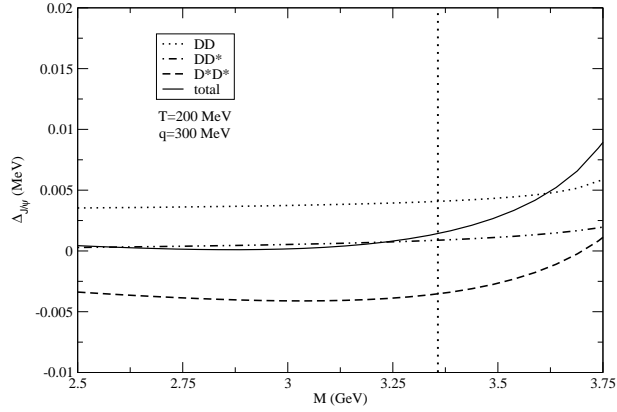}
\caption{Real part of $J/\psi$ self-energy for $D\oD$, $D\oD^*$, $D^*\oD^*$ 
loops.}
\label{Re_J}
\end{center}
\end{figure}

The above discussions indicate that the opening up of subthreshold decay
channels of $J/\Psi$ as shown here is primarily on account of the broadening
 of the spectral
shape of the charmed mesons and not so much due to the shift of their pole
positions towards lower masses. It is nevertheless worthwhile to investigate
the role of the effective mass of the $J/\Psi$. We thus evaluate its
self-energy following the same formalism (as for the $D^*$ meson) for $D\bar D$,
$DD^*$ and $D^*D^*$ loops. The spin averaged real part is given by
eq.~(\ref{real}) where the expressions for $N(q,k)$ for the three cases are
given by
\bea 
N^{D\bar D}(q,k)&=&\frac{2g_\psi^2F^2}{3}(q^2+4k^2-4k\cdot q)\nn\\
N^{DD^*}(q,k)&=&\frac{8g_\psi^2F^2}{3m_\psi^2}(q^2k^2-(k\cdot q)^2)\nn\\
N^{D^*D^*}(q,k)&=&\frac{2g_\psi^2F^2}{3}\left[18(k^2-k\cdot q+q^2)
+\frac{q^2}{m_D^4}(k^2q^2-(k\cdot q)^2)\right.\nn\\
&&-\frac{1}{m_D^2}(6k^4-12k^2k\cdot q+8(k\cdot q)^2)-2q^2k^2+3q^4)~.
\eea
In fig.~(\ref{Re_J}) we plot the mass shift of the $J/\Psi$,
$\Delta_\psi=\sqrt{m_\psi^2+{\rm
Re}\overline\Pi}-m_\psi$ as a function of its invariant mass at $T$=200 MeV. 
The shift turns out to be negligible at the pole.

\section{Summary and Conclusion}
In summary, motivated by the non-trivial drag and diffusion coefficients of $D$
mesons in hadronic matter seen in~\cite{Sabya_D}, \cite{Santosh_BD} and
\cite{Santosh_raa} we have performed a detailed analysis of
the spectral properties of heavy open charm mesons in hadronic matter at finite
temperature. We use the real-time formalism of thermal field theory and extract
the imaginary part of the self-energy from the discontinuities in the complex
energy plane. We obtain a small spectral broadening of the charm mesons
at finite temperature which enables the  
$J/\psi$ to decay into the subthreshold $D\oD$, $D^*\oD$ 
as well as $D^*\oD^*$ channels thus contributing to its suppression in
hadronic environment. We also find that the mass of the $J/\Psi$ does not change
appreciably in this model. The magnitude of the real part and the decay width
however depend crucially on the couplings as well as the cut-off employed.

{\bf Acknowledgment:} 
The authors thank B. K. Patra and G. Krein for useful suggestions.

\end{document}